\documentclass[nofootinbib,aps,10pt,twocolumn,pre,superscriptaddress]{revtex4-1}
\usepackage{rotate}
\usepackage{epsfig}
\usepackage{graphicx}%
\usepackage{psfrag}
\usepackage{tikz}
\usepackage{amssymb}
\usepackage{bm}
\usepackage{subfigure}

\newcommand \be  {\begin{equation}}  
\newcommand \bea {\begin{eqnarray} \nonumber }  
\newcommand \ee  {\end{equation}}  
\newcommand \eea {\end{eqnarray}}  

\begin{document}

\title{Can the glass transition be explained without a growing static length scale?}

\author{Ludovic Berthier}

\affiliation{Laboratoire Charles Coulomb (L2C),
University of Montpellier, CNRS, Montpellier, France}

\author{Giulio Biroli}

\affiliation{Institut de physique th\'eorique, Universit\'e Paris Saclay, CEA, CNRS, F-91191 Gif-sur-Yvette, France}

\affiliation{Laboratoire de Physique Statistique, \'Ecole Normale Sup\'erieure, CNRS, PSL Research University, Sorbonne Universit\'e, 75005 Paris, France}
 
\author{Jean-Philippe Bouchaud}

\affiliation{Capital Fund Management, 23 rue de l'Universit\'e, 75007 Paris, France}

\author{Gilles Tarjus}

\affiliation{LPTMC, CNRS-UMR 7600, Sorbonne Universit\'e, 4 Pl. Jussieu, 75005 Paris, France}

\begin{abstract}
It was recently discovered that SWAP, a Monte Carlo algorithm that involves the exchange of pairs of particles of differing diameters, can dramatically accelerate the equilibration of simulated supercooled liquids in regimes where the normal dynamics is glassy. This spectacular effect was subsequently interpreted as direct evidence against a static, cooperative explanation of the glass transition such as the one offered by the random first-order transition (RFOT) theory. We review several empirical facts that support the opposite view, namely, that a local mechanism cannot explain the glass transition phenomenology. We explain the speedup induced by SWAP within the framework of the RFOT theory. We suggest that the efficiency of SWAP stems from a postponed onset of glassy dynamics, which allows the efficient exploration of configuration space even in the regime where the physical dynamics is dominated by activated events across free-energy barriers. We describe this effect in terms of `crumbling metastability' and use the example of nucleation to illustrate the possibility of circumventing free-energy barriers of thermodynamic origin by a change of the local dynamical rules. 
\end{abstract}

\date{\today}

\maketitle

\section{Introduction}

What is the physical mechanism explaining the dramatic slowing down of glass-forming liquids as the temperature is decreased? This question has motivated a vast amount of work since the late 50's~\cite{review1,review2}. The seminal intuition of Adam and Gibbs is that atoms must move more and more collectively in order to flow~\cite{AG}, leading to an increase of the activation barrier as the glass transition temperature $T_g$ is approached. The idea of an underlying `amorphous order' that sets in over larger lengthscales has progressively been confirmed, as a result of intense theoretical~\cite{biroli-bouchaud,montanari}, experimental~\cite{Ladieu,Albert} and numerical efforts~\cite{cavagna1,cavagna2,glen,walter,sho,sho2,ceiling} in the last 20 years. It is now well-accepted that a static lengthscale grows, albeit modestly, in supercooled liquids approaching the glass transition. 

However, the physical relevance of these static correlations for the abrupt dynamical slowdown of supercooled liquids is still actively debated. The `elastic picture' for instance proposes that the chief physical ingredient driving the glass transition is the growth of the plateau shear modulus, $G_{\text{pl}}$, which makes even local moves progressively more difficult~\cite{dyre}. The growth of the activation barrier to flow would then simply mirror the growth of $G_{\text{pl}}$, without having to invoke any growing lengthscale. This purely local point of view was developed by Dyre \cite{dyre,dyre-review} and further promoted by Wyart \& Cates (WC) \cite{wyart-cates}, who take stock of the recent numerical results on the influence of particle swaps on the dynamics of polydisperse mixtures~\cite{hs2016,ninarello2017,ultrastable,misakijamming,daniele2018,2d}. The swap Monte Carlo algorithm (simply denoted here by SWAP) allows permutations of pairs of particles with different diameters~\cite{swap1,swap2,swap3,cavity1}. SWAP can be thought of as the introduction of an additional fluctuating degree of freedom attached to each particle --- its diameter~\cite{ninarello2017}. The physical dynamics is recovered when diameters are no longer allowed to fluctuate and only displacements of the particles are permitted. For well-chosen models, such a change in the {\it local} dynamical rules leads to a spectacular acceleration of the equilibration, reducing the relaxation time by several orders of magnitude~\cite{hs2016,ninarello2017}. 

In a nutshell, the WC argument is that if local rules are so important for the dynamics, then collective effects, while undisputably present, can only play a minor role in the $10^{15}$-fold increase of the relaxation time occurring when a liquid freezes into a glass. WC further argue that the violation of the Stokes-Einstein (SE) relation between viscosity and diffusion~\cite{ediger2000} can be used to bound from above the influence of collective effects in the slowing down of the dynamics. Since the violation factor at $T_g$ is about $10^3$ for fragile liquids --- compared to the $10^{15}$ increase of the relaxation time --- WC argue that 4/5 of the slowing down in log-scale should be attributed to local effects, thus disputing the experimental relevance of a static correlation length for glass formation. 

Three more papers have successively appeared on the same topic. Firstly, Ikeda {\it et al.} \cite{Ikeda} use a mean-field glass model introduced earlier by Mari and Kurchan \cite{mari-kurchan} to propose a theoretical description of SWAP. In this model, SWAP acceleration can be explained by a downward shift, computed by means of a static replica calculation, of the critical temperature of the mode-coupling transition once swap moves are allowed. Secondly, Brito {\it et al.} \cite{brito2018} used computer simulations and heuristic stability arguments to suggest that SWAP delays the onset of activated dynamics to lower temperatures than in the physical dynamics. Thirdly, a dynamic mode-coupling calculation was performed~\cite{szamel2018}, which shows that coupling diameter and density fluctuations can shift the location of the mode-coupling singularity to a lower temperature (or a higher density), while of course maintaining all static observables unaffected. 

Overall, these investigations provide sound explanations for the dynamic acceleration due to SWAP, with which we basically agree. However, they differ on the mechanism responsible for the slowing down of the dynamics when the glass transition is approached in the absence of SWAP, which in the end is the real question we are interested in. Is this mechanism purely local, as originally advocated by Dyre~\cite{dyre}? Is it `quasi-local', i.e., related to vibrational properties rather than static glassy correlations, as envisaged by Brito {\it et al.}~\cite{brito,brito2018}? Are instead collective rearrangements on the scale of the static correlations crucial to understand the physical nature of the glass transition? 

The aim of this paper is to revisit these questions, arguments and scenarios from the point of view of thermodynamic theories of the glass transition, and more specifically the Random First-Order Transition (RFOT) theory~\cite{rfot1,rfot2,rfot2b,rfot3}. 
Our two basic claims directly conflict with the local scenario summarized above:

\begin{enumerate}

\item The dramatic speedup of the dynamics induced by local changes in the dynamical rules is actually consistent with the RFOT theory, and more generally with static, cooperative explanations of the glass transition~\cite{AG,FLDT}. We describe this in terms of a `crumbling metastability', by which we mean that a metastability of collective thermodynamic origin can be postponed to lower temperature and higher pressure by purely local dynamical rules that leave thermodynamic properties unchanged. We provide a concrete illustration of this phenomenon in the case of crystal nucleation.  

\item The Stokes-Einstein decoupling can remain mild in the RFOT scenario, because local permutation processes are prohibitively more costly than collective relaxation. In this case, WC's upper bound on the influence of collective relaxation is not apposite. 

\end{enumerate}

The paper is organised as follows. 
In Sec.~\ref{rfot} we provide a short recap of the RFOT theory of glass formation. 
In Sec.~\ref{crumbling} we discuss metastability in finite dimensions and show that the change of local dynamical rules can allow a supercooled liquid to get around a barrier of thermodynamic origin and lose its metastability.
In Sec.~\ref{swap-rfot} we discuss how to best interpret SWAP efficiency within the RFOT framework.
In Sec.~\ref{local-collective} we critically assess the ability of local mechanisms to account for the phenomenology of glassy liquids.
We conclude our paper in Sec.~\ref{conclusion}. 
We also discuss kinetically constrained models and the Mari-Kurchan model in Appendix~\ref{app:other}, and present a simple bootstrap approach to emerging rigidity in Appendix~\ref{sec:bootstrap}.

\section{RFOT theory: A short recap}

\label{rfot}

The RFOT theory of glasses~\cite{rfot1,rfot2,rfot2b,rfot3} is inspired by mean-field spin-glass models in which the analogue of the liquid state becomes rigid in a two-step process as the temperature is lowered~\cite{wolynes1,KT,MP}. At a first temperature $T^*$, some incipient local rigidity, absent for $T > T^*$, allows metastable states to appear and `trap' the system for some amount of time. These metastable states are exponentially numerous, with an associated positive configurational entropy, $\Sigma(T)$. The mere existence of such a large number of metastable states allows the system to decorrelate with time: it is still a liquid, albeit one with some transient rigidity described by a nonzero shear modulus $G_{\text{pl}}$. At a lower temperature $T_K < T^*$, the configurational entropy vanishes, and the system undergoes a phase transition to an ideal glass phase. These statements can be made sharp in mean-field situations. However, their interpretation for realistic finite-dimensional systems is delicate and, although enticing, is still under construction. 

The temperature $T^*$ separating an essentially free-flow, liquid regime from an activated regime whose existence was conjectured long ago by Goldstein~\cite{goldstein}, corresponds to the modern interpretation of the Mode-Coupling transition (MCT) temperature~\cite{MCT}. In mean-field theory, this is a genuine dynamical phase transition which is accompanied by the divergence of a dynamical correlation length $\xi_{\rm dyn}$~\cite{MCTlength} and of thermodynamic barriers between metastable states. However, it is not clear what remains of this transition in finite dimensions as fluctuations appear to play a major role \cite{rfot3}. Similarly, below $T^*$ when the configurational entropy is presumably nonzero, metastability cannot be consistently defined beyond some finite `point-to-set' static length scale $\ell_{\text{ps}}$ (that diverges when $\Sigma(T) \to 0$)~\cite{biroli-bouchaud}. Since $\ell_{\text{ps}}$ is expected to be small close to $T^*$, the divergence of $\xi_{\rm dyn}$ is presumably cut off prematurely~\cite{sandalo}, leading to a mere crossover to a locally rigid state for $T < T^*$, characterized by a plateau in the relaxation function which is absent at higher temperature. This two-step relaxation is one of the landmarks of the MCT phenomenology~\cite{MCT}. 

The mode-coupling crossover at $T^*$ plays a crucial role in the understanding of how SWAP may speedup the thermalization of supercooled liquids. In fact, all recent explanations put forward in the literature \cite{Ikeda,brito2018,szamel2018}, including WC's original one, rely on a shift of $T^*$ to a lower temperature $T^*_{\text{swap}}$ due to SWAP dynamics. Within mean-field theory the temperature $T^*$ can be detected by analyzing a replicated system in the formalism of Franz and Parisi~\cite{FP}. Interestingly, it is numerically found that the emergence of a nontrivial Franz-Parisi potential in three-dimensional glass-formers takes place close to the MCT transition~\cite{simuFP1,simuFP2,ceiling}. Physically, this also signals the emergence of nontrivial static properties, which in mean-field theory corresponds to the appearance of metastable states~\cite{FP,charlotte}.  

When $T$ is further reduced below $T^*$, one enters the so-called mosaic state~\cite{rfot1}, where locally rigid, frozen regions of size $\ell_{\text{ps}}$ have to relax collectively for the system to flow. Within the RFOT theory, the associated energy barrier $B_{\text{coll}}$ to such collective rearrangements grows as 
\be
\label{eq:bcoll}
B_{\text{coll}}(T) = \Delta(T) \left[\ell_{\text{ps}}(T)\right]^\psi,
\ee
where $\Delta(T)$ is an energy scale, $\psi$ a certain exponent and $\ell_{\text{ps}}$ is argued to grow as the temperature is decreased.
The corresponding growing free-energy barrier is, very much in the spirit of the classical Adam-Gibbs mechanism, the explanation for the strongly non-Arrhenius, Vogel-Fulcher-type increase of the relaxation time in fragile liquids. It is important to emphasize that $\Delta(T)$ is nonzero only when the system is locally rigid, i.e., when $G_{\text{pl}}(T) > 0$, which occurs below $T^*$. Local stability is clearly a prerequisite to activation. 

\section{Crumbling metastability}
\label{crumbling}

\subsection{Metastability in finite dimensions}

All of the above theoretical construct relies on a liberally-used but rather elusive concept, that of `metastable states'. In fact, the main theoretical difficulty posed by the glass transition precisely lies in correctly handling this concept in non-mean-field situations.

Metastability is intrinsically a dynamical property. The fact that a collection of micro-states forms a {\it bona fide} metastable state depends both on the dynamical rules and on a timescale. This timescale should allow the system to evolve among the given set of micro-states (which define the metastable state) and yet be short enough for not allowing escape from the latter \cite{birolikurchan,schulman,bovier}. Such a separation of timescales can hold for one set of dynamical rules and not for another. 

In many physical systems, the mechanism leading to the existence of metastable states is of thermodynamic origin, as, e.g., for supercooled liquids which are metastable with respect to the crystal, superheated liquids which are metastable with respect to the gas, etc.~\cite{debenedetti,cavagna09}. In the case of glass-forming liquids the situation is more intricate since the role of thermodynamics is still hotly debated. Purely kinetic effects (i.e., dynamical rules) may play a dominant role in inducing metastability, even when the thermodynamic landscape is trivial. This is the idea put forward by theories of the glass transition based on kinetically constrained models~\cite{reviewKCM,garrahan-chandler,steve}. One then expects that if changing the local dynamical rules removes some of the constraints, this type of metastability will be destroyed (see also Appendix~\ref{app:KCM}). This is somehow the core of the argument used by WC (and also Ikeda {\it et al.}) to explain the acceleration generated by the SWAP, in which the strict enforcement of fixed particle diameters in a polydisperse mixture is waived. 

The main point we want to make is that even in cases where metastability is of thermodynamic origin, the temperature (or density) at which metastable states emerge and govern the physics of the system may depend on the local dynamics. Thus, by changing the local dynamical rules, the onset of activated glassy dynamics can be postponed to lower temperature or higher pressure.

\subsection{Affecting thermodynamic barriers by changing local dynamics}

While it is clear that a change in the local dynamical rules can allow the system to navigate in configuration space by avoiding {\it some} barriers, the question of whether such a change lowers or circumvents {\it thermodynamic} barriers is more subtle. By thermodynamic barriers, we mean free-energy barriers resulting from a thermodynamic drive involving collective behavior over a typical lengthscale that is definable through static observables only; and the considered change of dynamics is `local' in that the allowed elementary moves always involve a limited number of atoms that does not grow as one changes the control parameter(s), unlike the static length.  

WC's main point is that thermodynamic barriers cannot be altered by a change in local dynamics and, hence, observing such a change necessarily invalidates to a large extent a thermodynamic description. While this is likely true for asymptotically large barriers, such a general statement has limited scope in practice, as we now show in the well-understood case of crystal nucleation.

Below the melting temperature a supercooled liquid is metastable with respect to the crystal. It is known that the possibility to avoid crystallization and subsequently to form a glass involves kinetic effects, as illustrated by the form of the so-called TTT (Time Temperature Transformation) diagrams~\cite{debenedetti}. However, in this case the kinetic part is usually associated with the growth process and is {\it a priori} unrelated to the thermodynamic component of the time for crystallization given by the rate of nucleation $r_{\rm nucl}$. One can express the time to crystallization as the product
\begin{equation}
\tau_{\rm xtal}\propto \tau_{\rm kin}\, r_{\rm nucl}^{-1}\, .
\label{eq_nucleation}
\end{equation}
It is well established that: (1) $\tau_{\rm kin}$ is slaved to the typical relaxation time in the liquid\footnote{In practice, one may argue that it is rather the diffusion time that should be considered  but the difference is immaterial to our argument, as we will study a moderately supercooled liquid in which the decoupling between diffusion and $\alpha$-relaxation is not significantly large.}; (2) the nucleation rate, $r_{\rm nucl}$, is of {\it thermodynamic origin}   
and goes as $\exp(-\Delta F/k_B T)$, with $\Delta F$ the free-energy of the critical nucleation droplet; $\Delta F$ remains finite in the thermodynamic limit~\cite{debenedetti}.

To demonstrate that a change of the local dynamical rules can in some cases alter the thermodynamic barrier $\Delta F$ to crystallization, and not only the kinetic contribution, we have studied crystal formation in one of the three-dimensional supercooled polydisperse mixtures for which SWAP has proven to be very efficient \cite{ceiling,ninarello2017}. We compare results obtained with the standard Monte Carlo dynamics and with that in which SWAP moves are allowed. Our first observation is that SWAP is often so efficient that many putative polydisperse glass-forming liquid models crystallize even before being able to thermalize in the metastable supercooled liquid phase \cite{ninarello2017}. This, in itself, is already an indication that metastability and nucleation phenomena can be circumvented by a change of the local dynamics.

In order to study this effect in more detail, we have chosen an $11\%$ polydisperse mixture of hard spheres in the moderately supercooled (or rather supercompressed) phase, at pressures that are just above the melting one, and followed the time evolution of several samples by constant pressure Monte Carlo simulation with and without swap moves. The complex crystalline structures formed by a system very similar to the present one were recently analysed thanks to SWAP~\cite{truskett}.

\begin{figure}
\includegraphics[width=8.5cm]{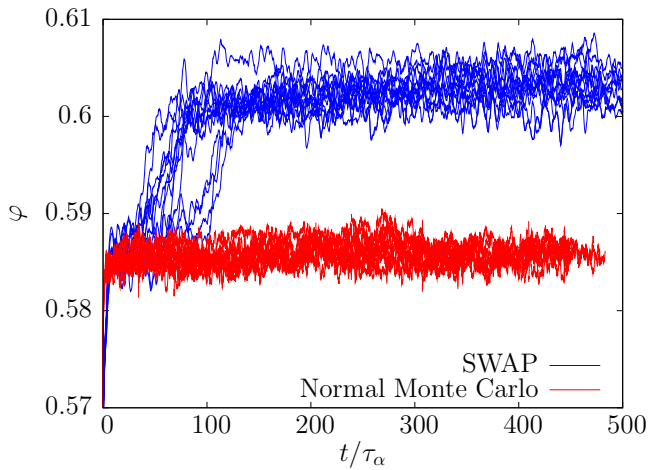}
\caption{Time evolution of the packing fraction $\varphi$ for a  three-dimensional polydisperse hard-sphere model following a pressure quench from $P=14$ (where $\varphi_{\rm eq} \approx 0.568$) to $P=16.2$ (where $\varphi_{\rm eq} \approx 0.586$). Ten independent samples of $N=1000$ particles are studied by SWAP and ordinary Monte Carlo simulations. The time $t$ is measured {\it relative} to $\tau_\alpha$ extracted separately for the two algorithms. For all samples, the system evolved under the ordinary dynamics remains in the metastable liquid phase up to the longest simulated time. On the other hand, the same samples evolved with SWAP started to crystallize within 10-70~$\tau_\alpha$ and metastability can barely be defined operationally.}
\label{xtal}
\end{figure}

The most interesting outcome of our computer study is displayed in Fig.~\ref{xtal} where we show the packing fraction after a quench as a function of time measured {\it relative} to the $\alpha$-relaxation time of the liquid, as obtained from the appropriate dynamics with or without swap. We observe that over the time span that we have been able to cover (which is of the order of $450\tau_\alpha$ at $P=16.2$), all of the samples when evolved with the ordinary dynamics have remained in the metastable liquid phase, whereas the same samples evolved with SWAP have all crystallized extremely fast. Since according to Eq.~(\ref{eq_nucleation}) the crystallization time divided by $\tau_\alpha$ is directly related to the thermodynamic barrier encountered by the system, our results show that the finite nucleation barrier thwarting crystallization in the ordinary dynamics is reduced or bypassed by introducing swap moves, which leads to a greatly accelerated nucleation process (on top of the expected effect on the kinetic term $\tau_{\rm kin}$)\footnote{A related effect was reported in an earlier work~\cite{sanz2007} where it had been shown that introducing swap moves could change the path to crystallization in binary mixtures when distinct forms of crystals can form during nucleation.}.

In fact it is even difficult to get an accurate measure of $\tau_\alpha$ using SWAP, since the system crystallizes over a timescale comparable to the equilibration time (i.e., before the correlation function approaches zero). By contrast, we have not been able to crystallize the system without SWAP  at any pressure, even when using many independent samples that were run over extremely large times (in units of $\tau_\alpha$). This directly shows that a well-established metastability for crystal nucleation --- which is of thermodynamic origin --- can totally `crumble' as a result of purely local changes in the dynamics. The observed speedup might be due to a reduction of the standard nucleation barrier with a critical nucleus similar to the non-SWAP case, or to the opening of a completely different channel bypassing that barrier. This is in itself an interesting question that requires further investigations, both in the crystallisation case and in the more complex glassy relaxation case.

The phenomenon in which metastable states are destabilized by a change in the local dynamics can therefore take place whether barriers are of thermodynamic origin or not (provided they are not too large, see next subsection). We expect this crumbling metastability to be rather dramatic when considering the complex free-energy landscape of glass-forming liquids. Before building on this aspect to discuss SWAP efficiency within the context of the RFOT theory in Sec.~\ref{swap-rfot}, it is useful to discuss in more detail different crossover temperatures relevant for nucleation, which have counterparts in the case of glassy dynamics as well -- although a strict analogy with nucleation can be misleading in that case.

\subsection{Two characteristic temperatures}
\label{two-temp}

For the sake of concreteness, we henceforth consider temperature as the control parameter, but depending on the system and physical conditions, the control parameter may instead be pressure, applied magnetic field or chemical potential, density, etc. Although our discussion applies to all cases in which metastability is of thermodynamic origin, it is convenient to keep the example of nucleation associated with a conventional first-order transition, as the above crystallization case or the Ising model in dimension $d\geq 2$. 

Metastability requires the existence of different thermodynamic states that can be envisaged as coexisting under some conditions. In finite-dimensional systems, nucleation is a mechanism through which, when changing the control parameter(s), the less stable states disappear and transform into the most stable one, and this activated process is driven by a thermodynamic force. All of this can only exist below a certain temperature $T_{\rm onset}$, which in the Ising model or the liquid-gas transition of a fluid is the critical temperature $T_c$. However, the temperature $T^*$ below which one can indeed observe metastability is generically lower than this upper limit of metastability. In the example of the Ising model, the metastability of a negatively magnetized state in the presence of a positive magnetic field $H$ appears at a temperature $T^*$ strictly below $T_{\rm onset}=T_c$, and it depends on the magnetic field: $T^*(H) < T_{\rm onset}$. Whereas the temperature $T_{\rm onset}$ is only determined by the thermodynamics of the system, $T^*(H)$, which signals a crossover at which the effect of metastability can be observed (i.e. when the rate $r_{\rm nucl}$ becomes small), depends on the local dynamics and on the observation timescale. 

Purely kinetic effects can only alter thermodynamic metastability under some conditions, and one expects that one such condition is that the lengthscale characterizing the nucleation process and the resulting escape from a metastable state are not too large. As a matter of fact, in the case of the Ising model, it can be proven that for small magnetic fields, when the nucleation size is very large, the nucleation barrier is independent of local dynamics and given by the thermodynamic nucleation argument \cite{martinelli}. Yet, in cases where characteristic lengthscales are not large, a change in the local kinetic rules can have a dramatic effect and destroy metastability, as we have illustrated above in the case of crystallization.

\section{SWAP efficiency within RFOT theory}
\label{swap-rfot}

\subsection{Characteristic temperatures/densities}

A major tenet of the RFOT theory is that activated dynamics in supercooled liquids is due to the emergence of amorphous order and cooperative rearrangements over a lengthscale $\ell_{\text{ps}}$. Within mean-field theory (more precisely within Kac models~\cite{kac}) this happens below a well-defined temperature $T^*$ corresponding to the point at which the equilibrium thermodynamic measure is fractured into many different basins. In this limit the {\it thermodynamic} barriers between these basins (or metastable states) diverge with the system size. As a consequence, $T^*$ is indeed independent of the local dynamical rules  and can be computed by a thermodynamic analysis without any reference to the implemented dynamics \cite{monasson,montanarimezard}. The mean-field treatment of SWAP dynamics developed by Ikeda {\it et al.}~\cite{Ikeda} is a smart approximation that accounts for kinetic effects within a purely thermodynamic computation but that is not representative of the standard behavior of finite-connectivity mean-field models with nonsingular interactions on which the RFOT theory is based and for which $T^*$ is unique -- see Appendix~\ref{app:MK} for an extended discussion of this point.

Within mean-field theory there also exists an onset temperature $T_{\rm onset}$, higher than $T^*$, which corresponds to the point at which the Franz-Parisi potential becomes nonconvex. In Kac models and in finite-dimensional glass-formers, $T_{\rm onset}$ can be identified with the temperature below which there exists a thermodynamic drive to metastability and is located where the point-to-set correlation length $\ell_{\text{ps}}$ starts to grow, i.e., when amorphous boundary conditions can stabilize metastable states in cavities of size $\ell_{\text{ps}}$. Around a temperature $T^* \leq T_{\rm onset}$, free-energy barriers are large enough to stabilize these metastable states for the considered dynamics. The temperature $T^*$ can be operationally defined as the empirical MCT temperature --- obtained for instance through a power-law fit of the relaxation-time data. 

As already pointed out, $T_{\rm onset}$ is independent of the microscopic dynamics whereas $T^*$ {\it a priori} depends on it. We will keep the notation $T^*$ for ordinary dynamics and call $T^*_{\text{swap}}$ the corresponding temperature for SWAP, and use similar definitions for the characterisitic densities $\varphi_{\rm onset}, \varphi^*$, and $\varphi^*_{\text{swap}}$. As we have shown in the previous section, SWAP allows a much faster equilibration in the example of the nucleation of the crystal phase in a metastable supercooled liquid, where barriers are of thermodynamic origin. Similarly, in glass-forming liquids around and below $T^*$ (or $\varphi^*$), SWAP is expected to wash out the metastability associated with the incipient amorphous order and to have dramatic consequences on the relaxation time in the range $T^*_{\text{swap}} < T < T^*$ (resp., $\varphi^*_{\text{swap}} > \varphi > \varphi^*$), as also suggested in Refs.~\cite{Ikeda,brito2018,szamel2018}. We illustrate this point below in the case of polydisperse hard-sphere systems.

\subsection{SWAP and effective landscape}

In equilibrium, the typical configurations visited by the system do not depend on the dynamical rules, provided detailed balance is satisfied. However, the `effective landscape' seen by the system, or more precisely by its representative point in configurational space, depends on the local dynamical rules: some channels allowing it to go from one configuration to another may be open for one type of dynamics and closed for another type. This is clearly true when one allows swap moves between particles of different diameters in a Monte Carlo algorithm. 
These moves correspond to displacements in real space that are at least of
the order of one particle diameter. (They also correspond to large
displacements in the configurational space associated with fixed-diameter
particles.) Introducing swaps is equivalent to
providing an extra dimension in which exchange of particles is much easier, while still fulfilling detailed balance. The landscape may also depend on the probability $p$ controlling the frequency of the swap moves in the Monte Carlo simulation. One can alternatively think of these moves as allowing each particle to change its diameter, which then adds one degree of freedom to each particle, with the constraint that the distribution of diameters is conserved at each step\footnote{This strict constraint can be softened by considering as in Refs.~\cite{brito2018,szamel2018} a grand-canonical version in which diameters are allowed to fluctuate, subject to a diameter-dependent chemical potential that enforces a constraint on the average size distribution only.}. This additional degree of freedom can be discrete or continuous depending on the exact nature of the polydisperse mixture.

\begin{figure}
\hspace*{-0.3cm}\includegraphics[width=8.4cm]{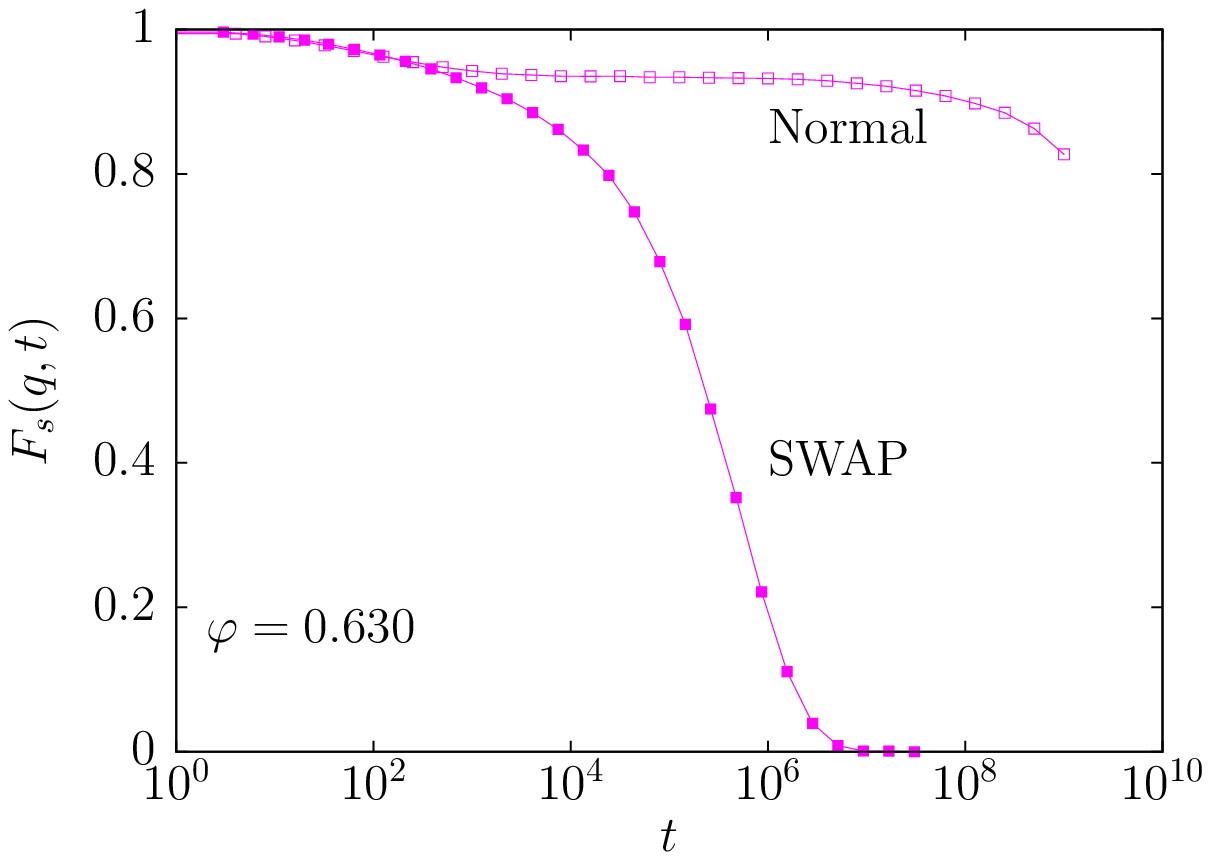}
\includegraphics[width=8.2cm]{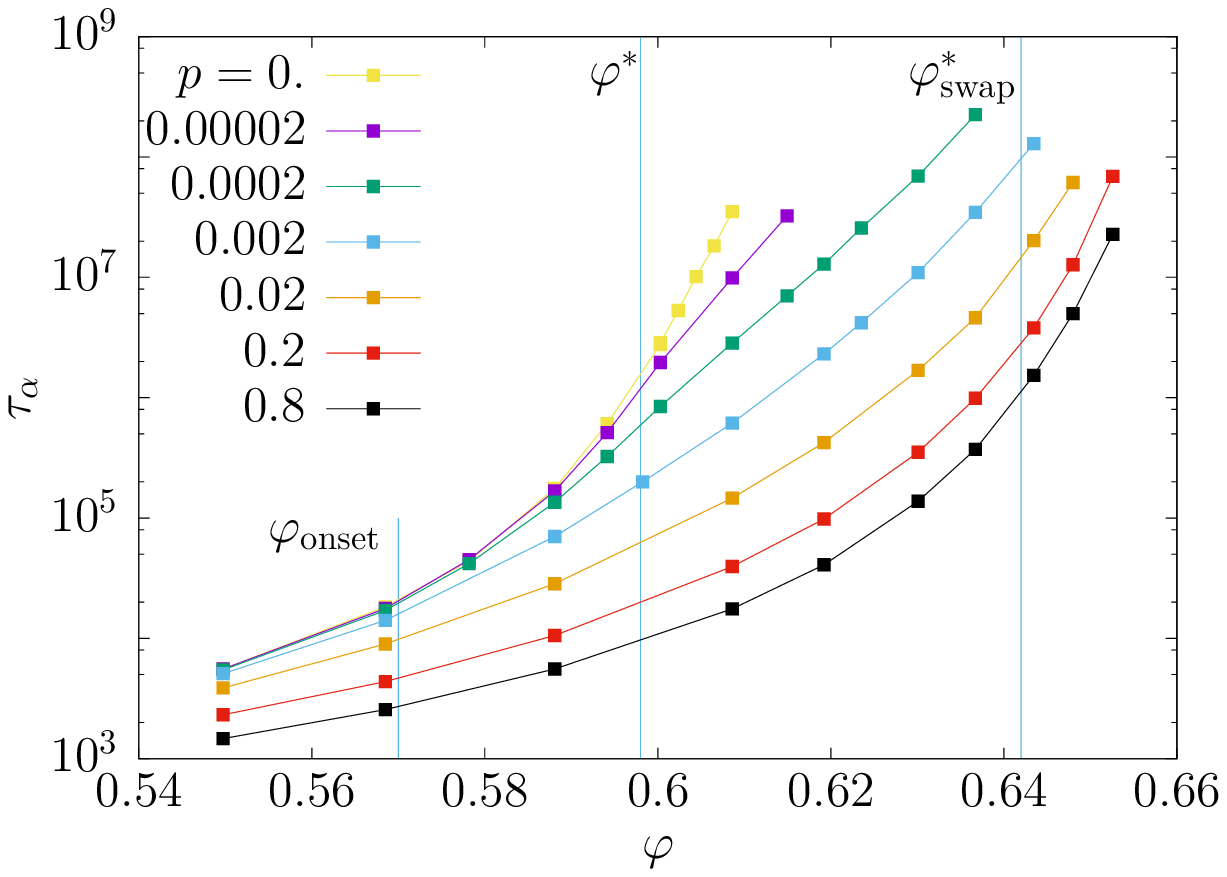}
\caption{Dynamics of a three-dimensional hard-sphere polydisperse model with and without SWAP.
Top: Self-intermediate scattering function at a packing fraction intermediate between $\varphi^*$ and $\varphi^*_{\rm swap}$, which illustrates the idea of crumbling metastability: The plateau corresponding to local metastability has completely disappeared with SWAP. $\varphi^*$ and $\varphi^*_{\rm swap}$ are empirically evaluated through an MCT power-law fit to the relaxation time data.
Bottom: Evolution of the equilibration time $\tau_\alpha$ with packing fraction $\varphi$ in SWAP Monte Carlo simulations where the fraction of swap moves, $p$, is varied between $p=0$ (ordinary dynamics) and $p=0.8$ (full swap dynamics). For intermediate $p$ values the dynamics smoothly interpolates between these two limits. SWAP relaxation for $p\geq 0.2$ appears more fragile than the normal Monte Carlo dynamics at $p=0$ (note the increased slope of the rightmost data points).}
\label{2landscapes}
\end{figure}

Opening new dynamical paths, as SWAP does, can only increase the number of unstable modes around stationary points and potentially destabilize states that would be metastable with the conventional dynamics. The effective landscape of the SWAP, which is a representation of configurations grouped into basins considered as metastable states over a given timescale and of the set of paths that connect one to another, must therefore be different from that in the absence of swap moves\footnote{The zero-temperature version of these free-energy landscapes are discussed in Ref.~\cite{brito2018} in the context of the jamming transition.}. 

As a case in point, we show in Fig.~\ref{2landscapes} the time dependence of the self-intermediate scattering function for the same three-dimensional hard-sphere model as studied in Refs.~\cite{hs2016,ceiling,ninarello2017}.  (The polydispersity here is 23~\% and is thus larger than the one used in Fig.~\ref{xtal}). We choose a volume fraction that is intermediate between $\varphi^*$ and $\varphi^*_{\rm swap}$. Strikingly, whereas the familiar two-step relaxation process is observed in the standard case, there is no longer any hint of an inflexion point in the swap case. Configurations that were metastable in the former case are completely unstable in the latter case: Locally rigid systems are fluidized by SWAP, i.e., glassy metastability has crumbled. A similar point was made in Ref.~\cite{brito2018} and is further discussed in Appendix~\ref{sec:bootstrap}.

The effective landscape being of dynamical nature, it depends in principle on the probability $p$ controlling the frequency of the swap moves compared to the translational ones in the simulation. This raises the possibility to explore a continuous range of effective landscapes, obtained by a continuous change of the probability $p$. We illustrate in Fig.~\ref{2landscapes} the effect of changing $p$ on the slowdown of relaxation of the same three-dimensional hard-sphere model as above. For this model, the empirically determined mode-coupling crossover occurs at a packing fraction $\varphi^* \approx 0.6$ and one can barely go beyond it for $p=0$. SWAP is most efficient with $p\approx 0.2$ \cite{hs2016,ceiling,ninarello2017}, because the dynamical speedup obtained by increasing $p$ further is not enough to compensate the increasing computational cost of the swap moves. This, then, corresponds to the optimal combined annealing of the diameter changes and particle displacements.  

For very small values of $p$, one observes a crossover from the normal dynamics at moderate $\varphi$ to the optimal swap dynamics at large $\varphi$, suggesting that all these dynamics in fact smoothly interpolate between only two extreme cases. This implies that as soon as $p>0$, the system will ultimately explore the effective landscape where particle diameters are allowed to fluctuate. It is therefore convenient to focus only on the two extreme dynamics: the conventional one with $p=0$ and the one with the optimal choice of $p$ corresponding to the best joint annealing of all degrees of freedom. (Note that throughout this paper the term SWAP refers to the latter.)

\subsection{Self-consistent description of activated dynamics}
\label{sec:self-consistent}

Within an RFOT picture the physical interpretation of the acceleration of the dynamics induced by swap moves goes as follows. Consider the situation in which one studies the thermalization of particles confined in a cavity with frozen amorphous boundary conditions~\cite{biroli-bouchaud}. Below $T_{\rm onset}$ and when the cavity radius is less than $\ell_{\text{ps}}$, the liquid inside the cavity cannot explore configurations typical of the bulk --- i.e., it is frozen too, in the sense that only a small subset of configurations have a significant weight in the Boltzmann measure. Relaxation of the density field is only possible when the radius is of the order of, or larger than $\ell_{\text{ps}}$. 

The mechanism envisioned by the RFOT approach to describe relaxation in the bulk is a self-consistent version of the above situation. In a bulk supercooled liquid, the boundary of any spherical subsystem of size $\ell_{\text{ps}}$ is only frozen on the timescale needed for a cavity of size $\ell_{\text{ps}}$ to relax. This picture only makes sense if this timescale is large, which is the case for the normal dynamics when $T < T^*$, since a positive free-energy barrier $B_{\text{coll}}(T)$, given by Eq. (\ref{eq:bcoll}), must be overcome. However, SWAP dynamics leads to larger dynamical fluctuations that effectively abolish $B_{\text{coll}}(T)$ in a temperature range $T^*_{\text{swap}} < T < T^*$.

Let us illustrate the above self-consistent argument by a simple model which suggests that metastability can indeed `crumble', i.e. vanish abruptly. When 
the environment of a typical subsystem of size $\ell_{\text{ps}}$ itself evolves with time, i.e., when it is not frozen forever, the barrier preventing the subsystem to relax is on average lowered (since the system will preferably relax when this barrier is exceptionally low). A toy phenomenological description of this effect is provided by
\begin{equation}
\label{eq_reduced_barrier}
B_{\text{coll}}(\tau) = B_\infty(T) \left[ 1-A \left(\frac{\tau_0}{\tau}\right)^a \right],
\end{equation}
where $\tau$ is the relaxation time of the surrounding, $\tau_0$ a microscopic time, $a > 0$ a phenomenological exponent, and $A$ a coefficient that depends on the microscopic dynamics and is larger in the presence of swap moves; $B_\infty\propto \ell_{\text{ps}}^\psi$ corresponds to the free-energy barrier {\it when the outside of the subsystem is frozen}, as in~\cite{biroli-bouchaud}. The precise form of the function describing the reduction of the barrier when $A$ increases or $\tau$ decreases is unimportant (the choice in Eq. (\ref{eq_reduced_barrier}) is for illustrative purpose only).

\begin{figure}
\psfig{file=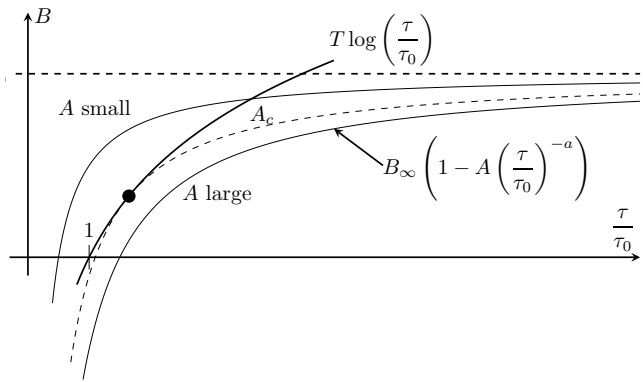,width=8.5cm,clip}
\caption{\label{fig:A} 
Graphical representation of Eq.~(\ref{crumpling}). We plot, as a function of $x=\tau/\tau_0$: $T \log x$ (thick plain line) and $B_\infty (1 - A x^{-a})$ for 3 values of $A$ (thin plain and dashed lines). For small values of $A$, there is one intersection point corresponding to the standard activated-time solution (with a barrier that scales as $B_\infty\propto \ell_{\text{ps}}^\psi$, yet with a dynamically renormalized prefactor). For large values of $A$, no intersection exists anymore, corresponding to crumbling metastability. The dashed line corresponds to the largest value of $A=A_c$ such that an intersection point exists. Note that $A_c$ increases when $B_\infty/T$ increases.}
\end{figure}

Now, the relaxation time of the typical subsystem of size $\ell_{\text{ps}}$ is self-consistently determined through the equation
\begin{equation}
\log \left( \frac{ \tau_\alpha}{\tau_0} \right) = 
\frac{B_{\text{coll}}(\tau_\alpha)}{T}.
\label{crumpling}
\end{equation}
It is easy to see graphically that this equation has an activated solution when $A$ is small enough, as we envisage for the normal dynamics: see Fig.~\ref{fig:A}. The relaxation time $\tau_\alpha$ decreases when $A$ increases, i.e., as the local dynamical fluctuations are increased, as in the case of SWAP. The activated solution then abruptly disappears for some value of $A$. This is our crumbling metastability scenario: Faster motion of the surroundings prevents the freezing of the inside of the cavity. Conversely, for a 
given $A$, an activated solution always appears at low enough temperature. 

The main effect of SWAP compared to the normal dynamics is therefore to delay the emergence of metastable states (and therefore of super-Arrhenius activated dynamics) from the temperature $T^*$ down to a lower temperature $T^*_{\rm swap}$, as also argued in~\cite{Ikeda}. Within the RFOT approach, the SWAP dynamics should slow down significantly near $T^*_{\rm swap}$, just as happens for the ordinary dynamics near and below  $T^*$. In this regime the SWAP dynamics should become super-Arrhenius activated, and one expects that asymptotically close to $T_K$ where $\ell_{\text{ps}}$ diverges, the (very large) barriers to overcome for relaxing the liquid should be the same with or without swap moves, as argued in Sec.~\ref{two-temp}. 

Now, since the temperature dependence of the relaxation time with SWAP is much weaker down to $T^*_{\rm swap}$ (and its accessible vicinity) and since the relaxation times for SWAP and ordinary dynamics must diverge in the same manner close to $T_K$, our picture suggests an extremely steep increase of the logarithm of the SWAP relaxation time in a narrower range of temperature $T_K < T < T^*_{\rm swap}$ than for the normal dynamics. This would correspond to an unusually `fragile' behavior. It is however likely that this takes place in a range which is difficult to equilibrate for SWAP itself, and that the convergence of the SWAP and non-SWAP relaxation times only occurs at astronomically long times. Still, the available data such as the one shown in Fig.~\ref{2landscapes} is not incompatible with this view. Indeed the density dependence of the normal Monte Carlo dynamics appears less sharp than the one of the swap dynamics with a large frequency of swap moves.

Although the interpretation put forward above uses similar ideas as in Wyart \& Cates~\cite{wyart-cates} and Brito {\it et al.}~\cite{brito2018} concerning the different effective landscapes for SWAP and normal dynamics, it differs completely on the conclusions one should infer about the physical dynamics. We strongly disagree with the claim that the SWAP efficiency disproves the relevance of activated processes over free-energy barriers controlled by thermodynamic properties for the normal dynamics near $T_g$, a claim that we find unsubstantiated. 

The disagreement might be related to the following subtle (and possibly confusing) considerations: 
\begin{itemize}
\item Whereas the equilibrium point-to-set length $\ell_{\text{ps}}$ is defined without any reference to the dynamics, it plays -- together with the local kinetic rules -- a crucial role in determining the relaxation time of the system. The reason is that only when $\ell \gtrsim \ell_{\text{ps}}$ is there an exponential (in $\ell$) number of escape paths available to the system, which allow decorrelation. But depending on the dynamics, only some of these paths are relevant, while others are too costly (or even forbidden) for kinetic reasons\footnote{Note that the kinetic constraints can be so strong, like in Kinetically Constrained Models, that all paths available on scale $\ell_{\text{ps}}$ are blocked and the system has to use even larger collective rearrangements, unrelated to thermodynamics. In this case, $\ell_{\text{ps}}$ becomes irrelevant for the dynamics.}. The point-to-set length $\ell_{\text{ps}}$ is not modified by swap moves, but, as argued above, its influence on the relaxation dynamics in the presence of SWAP becomes negligible -- at least when $B_\infty$ is not too large. 
\item The fact that the activated time needed to explore these escape paths grows exponentially with $\ell_{\text{ps}}$ does {\it not} require the system to be fully equilibrated within the cavity. Here, the analogy with standard nucleation is misleading: As emphasized in \cite{biroli-bouchaud}, $\ell_{\text{ps}}$ is not a critical nucleation radius (nothing nucleates at all within RFOT), but rather the first lengthscale for which activated events allow the system to explore more than a few basins in configuration space.
\end{itemize} 

\section{Local versus collective scenario for glass formation}
\label{local-collective}
 
\subsection{Local permutations and the Stokes-Einstein decoupling}
\label{SE}

Core to the argument of WC that collective effects play a minor role in the slowing down of relaxation leading to glass formation is the value of the Stokes-Einstein product between the self-diffusion constant and the viscosity estimated at $T_g$: $\mathcal{S}_g \equiv D(T_g) \eta(T_g)$. It is of 
order $\mathcal{S}_g \sim 10^3$ in fragile liquids. In their paper, WC suppose that the temperature-dependent barriers to relaxation in supercooled liquids add up as
\be \label{eq_B}
B_{\text{tot}}(T) = E_{\text{loc}}(T) + B_{\text{coll}}(T),
\ee
where $E_{\text{loc}}(T)$ is due to local barriers and $B_{\text{coll}}(T)$ is the collective contribution that can be related to the growth of the point-to-set correlation length $\ell_{\text{ps}}$. The idea put forward by WC is that, within RFOT, small-scale moves (on lengthscales $\ell < \ell_{\text{ps}}$) are unable to decorrelate the density fluctuations but should lead to local permutations of the particles, and hence to diffusion. In order to relax the structure one needs instead collective arrangements of local moves and hence a barrier $B_{\text{coll}}(T)$. As a consequence, the only barrier to diffusion should be the local energy barrier $E_{\text{loc}}(T)$. Thus, following WC, one expects the following behavior:
\begin{equation} 
D  \propto e^{-E_{\text{loc}}/k_B T}, \qquad \eta \propto e^{B_{\text{tot}}/k_B T},
\end{equation}
from which the amplitude of the decoupling is deduced: 
\be
\mathcal{S} = D \eta \propto e^{B_{\text{coll}}(T)/k_B T}\,.
\ee
WC's conclusion is therefore that $B_{\text{coll}}(T_g)/k_B T_g$ is at most $\log 10^3$, when the total barrier to relaxation at $T_g$ is, by definition, $B(T_g)/k_B T_g \approx \log 10^{15}$. Hence, according to WC, the collective barrier only explains about $1/5$ of the slowing down (in number of decades) from high temperature to the glass transition, namely, 3 decades out of the 15 decades observed for the viscosity increase. Taking into account dynamical heterogeneities would reduce even further the part of the decoupling due to collective barriers. 

There are several experimental and numerical facts that show that this scenario is moot. Very costly, local permutations must indeed become the dominant channel to single-particle dynamics very close to $T_K$ (because $B_{\text{coll}}$ then diverges whereas $E_{\text{loc}}$ remains finite), in which case we indeed expect a huge SE violation factor. However, both single-particle dynamics and collective-density relaxation use the {\it very same} collective mechanism when $T \gtrsim T_g$, and this also applies to diffusion.
Indeed, several studies of the SE factor have shown that the decoupling between diffusion and relaxation can be explained in terms of a spatial heterogeneity of the relaxation. More precisely, viscosity and diffusion probe different moments of the distribution of local relaxation times $\tau$~\cite{ediger2000,book}, i.e., $\eta \propto \langle \tau \rangle$ and $D \propto \langle \tau^{-1} \rangle$, which behave differently as temperature decreases, even when both these quantities are controlled by the same collective effects. Several direct indications that this picture is correct are: 
\begin{enumerate}
\item The amount of SE violation is empirically correlated with the exponent $\beta$ characterizing the stretching of the time-dependent relaxation function~\cite{ediger2000}, thus confirming the important role played by the broadness of the distribution of the local relaxation times; 
\item The same amount of decoupling between diffusion and structural relaxation is observed when considering exclusively single-particle dynamics (but over a range of wavevectors)~\cite{PRE2004}; in the WC scenario this should not happen since decoupling is a direct consequence of comparing a single-particle observable with a collective one;
\item The dynamical susceptibilities (four-point correlation functions) for collective and self density relaxation both show similar peaks for the same $\alpha$-relaxation time of the system \cite{glotzer}, thus suggesting that the {\it very same collective phenomenon} is responsible for both.
\end{enumerate}

The physical picture we envision for the dynamics on scale $\tau_\alpha$ is that each particle keeps more or less the same neighborhood, which rearranges as a whole when a collective event takes place. These collective events relax the density field but it is important to realize that they only allow the system to visit a subset of the possible equilibrium configurations. For instance, configurations corresponding to a single swap move are not available and need times much larger than $\tau_\alpha$ to be reached.\footnote{As emphasized at the end of the previous section, this however does not invalidate the arguments leading to an exponentially small trial frequency for events happening on scale $\ell_{\text{ps}}$.} 
The barriers to local permutations of particles are simply far too high to provide a competitive channel when $T \sim T_g$ (recall that we are now discussing the normal dynamics without SWAP). So in a sense we agree with WC that $E_{\text{loc}} \gg B_{\text{coll}}$ but draw from this observation the opposite conclusion. The collective channel is not in series with the very costly local permutation channel, as in Eq.~(\ref{eq_B}), but is in parallel with it. 

In a situation where local permutations are extremely costly (due to the strong repulsive interactions at short distance), the system will use the other excitation branch that involves collective motions on scales $\ell \lesssim \ell_{\text{ps}}$, for which the activation barrier is smaller than the typical value $B_{\text{coll}}(T)$. These events contribute to the distribution of local relaxation times mentioned above~\cite{wolynes-beta} and probably to the $\beta$-relaxation process as well, but they are not able to decorrelate typical density fluctuations and therefore do not contribute to the viscosity or the $\alpha$-relaxation time. 

Several phenomenological approaches have shown that such a description is able to reproduce the observed values of the SE factor $\mathcal{S}_g$ as well as other dynamical features of supercooled liquids \cite{viot2000,xia2001}. We therefore claim that the physical relaxation channel leading to diffusion and that leading to viscosity are in fact the same --- at least close to $T_g$ --- and has to be collective, i.e., has to involve a substantial number of particles. 

\subsection{Emerging rigidity: shortcomings of a `quasi-local' scenario}
\label{shortcoming}

Invoking a purely local explanation of the glass transition appears to us to be a step back to the pre-1995 state of affairs, before the flurry of activity around spatial heterogeneities in glass-forming liquids~\cite{ediger2000,book}. Below the crossover temperature $T^*$ where most theories envisage a change of nature of the relaxation in fragile supercooled liquids, a purely local scenario is clearly at odds with many experimental and numerical data. In particular it cannot be reconciled with the continuous growth of spatial correlations in the dynamics of supercooled liquids that is detected by multi-point space-time correlation functions~\cite{science,quedalle} and nonlinear responses~\cite{Albert,Ladieu}.

WC argue that the local scenario should rather be interpreted as `quasi-local' in the sense that local moves involve finite collections of particles \cite{wyart-cates}. The associated lengthscale is surmised to be the  dynamical correlation length $\xi_{\rm dyn}$, whose growth is aborted when $T \approx T^*$ (hence the quasi-local character below $T^*$). Following elastic models of viscous liquids such as the shoving model put forward by Dyre \cite{dyre,dyre-review}, WC implicitly assume  that
\be
E_{\text{loc}}(T) = G_{\text{pl}}(T) V_c,
\label{jeppe}
\ee
where $V_c \sim \xi_{\rm dyn}^3$ is constant below $T^*$ and $G_{\text{pl}}(T)$ is the plateau shear modulus.

However, WC~\cite{wyart-cates}, as well as Brito {\it et al.}~\cite{brito2018}, remain vague about the relaxation mechanisms around and below $T^*$ within their scenario, which nonetheless does not easily account for the well-established monotonous growth of multi-point correlation functions below $T^*$. In their view, spatial correlations in the dynamics should only be present near $T^*$, i.e., at the onset of rigidity where the soft modes associated to marginal stability become delocalized~\cite{brito}. Below $T^*$ the system departs from marginality and becomes increasingly more stable. This increased stability should naturally imply that the lengthscale related to the relaxation process [i.e., $V_c$ in Eq. (\ref{jeppe})] decreases, in contradiction with experimental and numerical results which show instead that this length continues to grow~\cite{quedalle}. The possibility, invoked by WC and Brito {\it et al.}, that the dynamics below $T^*$ is thermally activated along some soft dynamical modes spanning a length $\xi_{\rm dyn}$ that does {\it not} decrease is intriguing but lacks at present a substantial explanation, precise quantitative calculations and numerical support.

Another necessary input in a scenario based on Eq.~(\ref{jeppe}) is the emergence of a nonzero plateau shear modulus $G_{\text{pl}}(T)$ and its sizable increase of as one cools the system. Even if it is not precisely described in \cite{wyart-cates,brito2018}, the emergence of rigidity associated with the disappearance of soft marginal modes near $T^*$ appears akin to the MCT/RFOT scenario. As recalled in Appendix~\ref{sec:bootstrap}, a nonzero plateau shear modulus $G_{\text{pl}}(T)$ indeed appears rather abruptly and grows as $T$ is decreased below $T^*$, with a concave temperature dependence~\cite{yoshino2012}. Therefore, even with a nondecreasing $V_c$, Eq.~(\ref{jeppe}) would predict a concave (downward) temperature dependence of the activation barrier. This qualitatively disagrees with the empirical observation of an convex (upward) growth as temperature decreases (see~\cite{Tarjus2,rossler}), except if the experimental $G_{\text{pl}}(T)$ has itself a convex behavior, which would be at odds with most studies of the Debye-Waller factor (proportional to $T/G_{\text{pl}}$, see Appendix~\ref{sec:bootstrap}) in the vicinity of $T^*$.

Finally, some additional numerical and experimental results appear to favor the collective scenario over the WC scenario. For example, Larini {\it et al.}~\cite{Larini} have shown that for many glassy systems, one can find a master curve that collapses the dependence of the logarithm of the relaxation time $\log \tau$ as a function of the ratio $\langle u^2 \rangle_g/\langle u^2 \rangle$, where $\langle u^2 \rangle$ is the Debye-Waller amplitude of displacements in the plateau regime. According to Dyre \cite{dyre-review} and WC, this relation should be linear, while the master curve of Larini {\it et al.} reveals a large nonlinear component. Taking their functional fit seriously suggests that the nonlinear contribution to the increase of $\log \tau$ is 6 times larger than that of the linear contribution. This is qualitatively similar to the results of Buchenau {\it et al.}~\cite{Buchenau}, who conclude that the collective barrier contribution to fragility at $T_g$ is between 1 and 6 times that of the local barrier contribution, with a ratio that increases with fragility itself. These results again suggest that collective effects play a large role in the increase of the effective energy barrier $B_{\text{tot}}$ as the temperature is lowered.

\subsection{Insights from amorphous confinement}

All explanations of the dynamic acceleration due to SWAP make an assumption about the mechanism responsible for the dynamical slowdown in the absence of SWAP. These theoretical explanations must be consistent with experimental and numerical results obtained so far for supercooled liquids. Important physical 
facts are provided by  numerical results on confined liquids with amorphous boundary conditions.  Cavity measurements with frozen amorphous boundary conditions are performed in the first place to determine the static point-to-set correlation length $\ell_{\text{ps}}$. However, what is of further interest for the present discussion is the outcome of these studies for the equilibration dynamics. Two different timescales can be used: the relaxation time, i.e., the time it takes for the equilibrium correlation function inside the cavity to decorrelate, and the equilibration time, i.e., the time it takes for the system to reach equilibrium inside the cavity starting from a random initial condition (this was called the BIC test and used to check that simulations are indeed equilibrated in measurements of $\ell_{\text{ps}}$ \cite{cavity1}). Below we focus on the latter timescale.

The normal dynamics is highly sensitive to confinement and starts to slow down for cavity sizes that are significantly larger than $\ell_{\text{ps}}$~\cite{cavity1,cavity2}, presumably as large as the dynamic correlation length $\xi_{\rm dyn}$. The equilibration time increases dramatically as the cavity size is reduced further, to the point that measuring $\ell_{\text{ps}}$ is actually a prohibitively difficult problem~\cite{sho,cavity2}. SWAP dynamics is much more efficient than the normal dynamics in a frozen cavity as well, but SWAP itself dramatically slows down as the cavity size approaches $\ell_{\text{ps}}$~\cite{ceiling}. This effect, in itself, rules out to a large extent any purely local explanation of the slowing down of the normal dynamics, which once SWAP moves are allowed should be blind to $\ell_{\text{ps}}$.\footnote{Note that an interesting earlier attempt to understand how SWAP works in the context of the RFOT theory can be found in ~\cite{cavity1}. Our overall scenario substantially differs from the one proposed in this work.}

This dynamical phenomenon, which becomes more prominent as the point-to-set length increases, emerges below $T^*$ and above $T^*_{\rm swap}$. In this temperature regime, the bulk normal dynamics is deep into the (inaccessible) super-Arrhenius activated regime and the bulk SWAP dynamics is still very fast and characterized by a small dynamic correlation length $\xi_{\rm dyn, swap}$.

These facts have important implications. If the physical relaxation process were only governed by a local energy barrier $E_{\rm loc}$, SWAP, as already mentioned,  would fully bypass the glassy slowdown and would be completely insensitive to a confinement over a length controlled by the $\ell_{\text{ps}}$, contrary to numerical findings. In the `quasi-local' scenario put forward by Brito {\it et al.}~\cite{brito2018}, one should envision that the slowing down due to confinement has actually nothing to do with point-to-set correlations and rather takes place at the dynamical correlation length $\xi_{\rm dyn,swap}$ associated with $T^*_{\rm swap}$. However, from the simulation results this would mean that the dramatic effect of confinement on the dynamics kicks in when $\chi_4^{\rm swap}$ is still very small, a fact that is hardly justifiable especially in comparison with what happens for the usual dynamics.

As a final comment, which echoes those made at the end of section \ref{sec:self-consistent}, we stress that the behavior of the equilibration time inside a cavity as a function of $\ell$ is a proxy for the relaxation time of the bulk dynamics of a liquid on scale $\ell$. The facts discussed above show that in the studied temperature domain it is a strongly {\it decreasing} function of $\ell$ both for SWAP and normal dynamics. 
This provides direct support to what we put forward in the explanation of SE decoupling: Local equilibration in supercooled liquids is not achieved before the collective one, but at the same time. As discussed previously, the role of the point-to-set length on the $\ell$-dependence of the relaxation time, which differs depending on the dynamics, can be rationalized in terms of the explosion of the number of available escape paths at $\ell_{\text{ps}}$. Clearly, a detailed microscopic understanding of this behavior is crucial and future theoretical investigations should be devoted to it. 

\section{Conclusion}
\label{conclusion}

We have argued that the observations recently put forward to dismiss the role of collective effects in the dynamical slowing down of glass-forming liquids --- namely, the efficiency of the SWAP Monte Carlo algorithm to thermalize polydisperse glass-formers below $T_g$ and the Stokes-Einstein decoupling at $T_g$ --- can actually be accounted for within theories in which these collective effects are central. We contended that alternative local and quasi-local explanations of SWAP acceleration have trouble coping with several experimental and numerical results on supercooled liquids.  

Our main point is that introducing swap moves can wash out metastability and allow the system to bypass free-energy barriers, even of thermodynamic origin. This crumbling metastability postpones the temperature at which collective activated processes are required and therefore dramatically accelerate the dynamics of the system --- at least when the cooperatively rearranging regions are not too large. We have shown numerically that this is the case for standard nucleation. 

Several points argue against a local (or quasi-local) view of the dynamics, among which: (a) Diffusion and structural relaxation use the same collective channels in the normal dynamics, such that the rather mild Stokes-Einstein decoupling is a result of dynamical heterogeneities and therefore perfectly compatible with the standard RFOT picture, at least for temperatures around the glass transition; (b) the effective energy barrier defined as $k_B T \log (\tau_\alpha/\tau_0)$ is a convex function of $1/T$, while quasi-local theories with a saturating correlation length predict a concave dependence; (c) relaxation in a frozen cavity with swap dynamics slows down as soon as the size of the cavity reaches the point-to-set correlation length $\ell_{\text{ps}}$, whereas local (or quasi-local) barriers should be blind to $\ell_{\text{ps}}$.           

All this being said, one should keep in mind that $\ell_{\text{ps}}$ is never very large in real systems\footnote{In the case of the 3-dimensional polydisperse hard-sphere mixtures whose dynamics can be very efficiently accelerated by SWAP, the metastable liquid has been thermalized at densities larger than the expected laboratory glass transition~\cite{ceiling}. A point-to-set correlation length $\ell_{\text{ps}}$ can be detected above some $\varphi_{\rm onset}$ and the signature of nontrivial fluctuations of the overlaps is also found, all this in line with the predictions of the RFOT theory. However, the increase of $\ell_{\text{ps}}$ between $\varphi_{\rm onset}$ and the expected laboratory glass transition is only about a factor of $2.5$. This would still be sufficient to explain the slowdown of relaxation via an activated scaling formula but this increase is less than that found in other glass-forming models and less than expected for fragile molecular liquids (the growth being nonetheless modest in these cases). At this point, it is still unclear whether or not such polydisperse models present an intrinsic quantitative difference with other glass-forming liquids as far as their collective behavior is concerned.}, which is of course a consequence of the prediction of an activated dynamical scaling such as in Eq. (\ref{eq:bcoll}), but also opens the possibility that theories that seem at odds with the existence of a growing amorphous order could actually be somehow combined with it in a larger theoretical scheme. In fact, local elastic models and RFOT theory are more akin than may appear at first sight~\cite{wolynes-elast}. 

Our general conclusion is that the success of SWAP, while generating a genuinely useful debate about the validity of the RFOT approach, does not directly favor one view of the glass transition nor disproves any. However, since it allows the exploration of new and unexplored territory in glass physics~\cite{scalliet,gardner,deposited,yielding,lowTmodes}, we might expect some progress towards discriminating the relevance of various approaches, when their predictions are pushed and tested over an unprecedently wide range of temperature and studied as a function of dimension $d$ as well\footnote{Given that by lowering the dimension $d$,  fluctuations are expected to decrease thermodynamic barriers for a fixed value of the point-to-set length, we expect that SWAP should allow the exploration of a larger range of point-to-set lengths in two dimensions~\cite{2d}, and a smaller range in dimensions larger than three.}. We hope that progress along these lines will be available soon.

\acknowledgments
We thank M. Cates, D. Coslovich, R. Jack, J. Kurchan, F. Martinelli, P. G. Wolynes, M. Wyart, H. Yoshino, S. Yaida, and F. Zamponi for many useful discussions. This work is supported by a grant from the Simons Foundation (No. 454933, LB, No. 454935, GB).

\appendix

\section{Other scenarios for SWAP efficiency}
\label{app:other}

\subsection{Kinetically constrained models}
\label{app:KCM}

The dynamic facilitation approach to the glass transition~\cite{garrahan-chandler} mostly relies on the idea that dynamics slows down because of kinetic constraints which forbid some pathways~\cite{steve}. Barriers thus have a purely kinetic origin, and are totally unrelated to the thermodynamic properties of the system. The approach is illustrated by simple noninteracting spin models known as kinetically constrained models (KCM)~\cite{reviewKCM}. Although swap moves {\it per se} may not be relevant for such models, the success of a change of local dynamical rules seems easily explained conceptually, since the kinetic constraints are purely local and can thus, in principle, be extremely sensitive to the microscopic dynamics (see however below).  In a sense, it is an implicit prediction of this approach (although it was never truly put forward or explored) that certain types of local algorithms should be able to considerably accelerate the thermalization of these systems. Indeed in KCMs, thermalization of initial conditions is not a difficult issue at all, and only the dynamical relaxation processes are of physical interest.

On the other hand, this issue might be more subtle than it seems at first sight. In fact, in the only microscopic models where kinetic constraints were derived from first principles, the so-called plaquette models~\cite{juanpe}, any local change of the spin-flip dynamics would not be able to fully accelerate the dynamics. In these systems, dynamics is glassy at low temperature and there is a diverging static correlation length with a thermodynamic singularity at zero temperature. The origin of glassiness is the existence of dilute defects at low temperature. Local changes in the dynamics can only speed up the dynamics of defects by increasing their diffusion coefficient but not reducing their dilution. 

Ideally, one would like to know better how kinetic constraints emerge in the dynamics of realistic supercooled liquids, in order to invent smart Monte Carlo moves that at once preserve the equilibrium properties and bypass the kinetic constraints. 
For now, it is unclear why SWAP would achieve precisely this nontrivial goal. Similarly, it appears difficult to predict, in this view, what is the temperature dependence of the SWAP dynamics and what are its characteristics. Such an understanding would be useful, as it could allow the potential determination of even better algorithms to achieve thermalization at lower temperatures.  

\subsection{The Mari-Kurchan model}
\label{app:MK}

In their analysis of the Mari-Kurchan (MK) model~\cite{mari-kurchan}, Ikeda {\it et al.}~\cite{Ikeda} discuss the emergence of two distinct mode-coupling temperatures. Their main conclusion is that the normal dynamics essentially freezes at some temperature $T^*$, whereas the SWAP dynamics does at a lower temperature, $T^*_{\rm swap} < T^*$. Although this is along similar lines to ours and WC's discussion, the interpretation of some of these results and their meaning for finite-dimensional systems is quite different, as we now explain. 

Let us first address a crucial point. The MK model is a `mean-field' description of a finite-dimensional system but a very special one, where neither the MCT dynamical transition nor the thermodynamic glass transition are truly realized. Indeed, the genuine thermodynamic equilibrium of the model is attained when including all local moves, which comprise changes of particle diameters (the equivalent of the swaps) and individual particle hopping. In consequence, the emergence of a `high' transition temperature $T^*$ when arbitrarily preventing diameter changes results from a purely kinetic constraint. If instead one allows by {\it fiat} changes of diameters without changing the positions of the particles, freezing and lack of diffusion take place at a lower temperature, identified as $T^*_{\rm swap}$. If one allows all local moves, including diameter change and particle hopping, then no freezing takes place at any finite temperature. 

This situation is very different from generic finite-connectivity mean-field models. Indeed, for mean-field models on Bethe lattices where locality has a well-defined meaning, it is conjectured and to a large extent proven \cite{montanari} that any local change of the dynamics cannot generically alter the value of $T^*$. This statement is related to the general idea that infinite thermodynamic barriers, such as those emerging at $T^*$ in mean-field models, cannot be destroyed by any local change of the dynamics. For example, consider a finite temperature lattice-glass binary mixture on a Bethe lattice~\cite{bethe}. Such a model has a unique MCT transition, irrespective of the local dynamics (as long as it is an irreducible Markov chain), at which the Boltzmann measure breaks up into many thermodynamic states separated by infinite barriers. Certainly, the grand-canonical dynamics, which resembles the SWAP one, is faster than the canonical one but they both have the same $T^*$. This is the usual mean-field scenario on which the RFOT theory is based. In the zero-temperature limit, where the Markov chain becomes reducible since hard kinetic effects intervene, an MCT transition which is similar to those of KCM's on Bethe lattices \cite{bethe} can preempt the thermodynamic MCT transition. It is the counterpart of the former kind of transition that has been investigated by Ikeda {\it et al.} and that, arguably, can take place in other mean-field models as well, due to kinetic constraints in the dynamics\footnote{The infinite dimensional limit certainly plays an important role in inducing the kinetic transition found in \cite{Ikeda}.}.


 In short, the work of Ikeda {\it et al.} introduces a smart approximation to take into account kinetic effects within a static replica computation. We agree that it provides a possible explanation of the difference between $T^*$ and $T^*_{\rm swap}$, which is of kinetic origin. Overall, it reiterates from a mean-field perspective the dynamic facilitation view that kinetic constraints are the main cause of slow dynamics in supercooled liquids. Although this is certainly possible, and realized in some models endowed with specific dynamical rules, we think that experimental and simulation results for supercooled liquids with (standard) dynamics rather point toward an explanation in which a growing static length plays a key role.  
 
One of the main points of Ikeda {\it et al.} is that, depending on the local dynamics, one has to consider different kinds of overlap in the replica computation, which leads to a dynamical transition taking place at different temperatures. (As recalled above, it even disappears if all local dynamical moves are allowed.) If one combines this result with the usual definition of the point-to-set length, which is thermodynamical and thus includes all local dynamical moves inside a cavity, then the point-to-set length should start to increase only below $T^*_{\rm swap}$ in finite-dimensional systems. This is precisely where we disagree. As discussed in the main text, our view is that within the RFOT theory $T_{\rm onset}$ is the temperature at which there starts to be a thermodynamic drive toward metastability and at which, accordingly, the point-to-set length starts to grow. Numerical evidence suggests that this is indeed the case: The growth of $\ell_{\rm ps}$, together with the emergence of a nontrivial Franz-Parisi free energy, become observable in three-dimensional glass-formers between $T_{\rm onset}$ and the physical (i.e., without swaps) MCT crossover $T^*$, and not at the much lower temperature $T^*_{\rm swap}$.

In summary, our disagreement stems from the interpretation of the MK results and their implications to finite-dimensional supercooled liquids, not from the analytical results of Ikeda {\it et al.} which, we believe, are correct.

\section{A bootstrap theory for emerging rigidity}
\label{sec:bootstrap}

Clearly, the appearance of local rigidity must itself be a collective, bootstrap phenomenon. This idea permeates many approaches to the glass transition, from the ``cage'' picture advocated in the context of MCT to the isostatic theories of jamming. For a particle not to move easily, its neighbours must themselves be blocked, and so on. One can formalize this scenario by using the recent replica theory of Yoshino and  M\'ezard~\cite{yoshinomezard,yoshino2012} that allows one to estimate the plateau shear modulus $G_{\rm pl}$ from first principles. The physical idea is to write $G_{\rm pl}$ as a difference of two contributions
\begin{equation}\label{eq:born}
G_{\rm pl} = G_{\rm born} - G_{\rm fluc},
\end{equation}
where $G_{\rm born}$ is the so-called Born term coming from the contribution of affine displacements under shear, and $G_{\rm fluc}$ is a contribution induced by thermal fluctuations. In an ergodic, liquid, phase translation invariance imposes that $G_{\rm fluc} = G_{\rm born}$, and, thus, that the shear modulus is zero, as expected. At smaller temperatures, $G_{\rm fluc}$ decreases, allowing for the possibility of rigidity, at least on intermediate timescales. The detailed replica calculations are quite intricate~\cite{yoshinomezard,yoshino2012}, but one can grasp the correct scenario by approximating $G_{\rm fluc}$ as:
\begin{equation}\label{eq:fluct}
G_{\rm fluc} = \frac{b T}{G_{\rm pl}},
\end{equation}
where $b$ is a coefficient that depends on the microscopic details of the model. This equation states that local fluctuations in a solid are proportional to temperature and inversely proportional to the shear modulus (which assumes that the latter is small compared to the bulk modulus). Inserting Eq. (\ref{eq:fluct}) back into Eq. (\ref{eq:born}) leads to a second-order equation for $G_{\rm pl}$, whose solution is
\begin{equation}
G_{\rm pl} = \frac12 \left[G_{\rm born} + \sqrt{G_{\rm born} - 4 b T} \right]; \quad T < T^* \equiv \frac{G_{\rm born}}{4b}.
\end{equation}
This simple description predicts that the plateau modulus jumps from $0$ to a finite value at $T=T^*$, with a singular square-root contribution when $T < T^*$, exactly as predicted by standard MCT. Intuitively, MCT captures the self-consistent appearance of local rigidity, that can only exist if the rest of the system is itself rigid enough. By the same token, if the local dynamical rules (like SWAP) lead to an increase of local fluctuations (i.e., an increase of $a$), one expects $T^*$ to decrease. 
Note that one can rephrase the above argument in terms of bootstrap percolation with the effect of swaps modelled as an increased number of neighbors required to stabilize each particle~\cite{FranzSellitto,BiroliIkedaMiyazaki}. This latter formulation is conceptually very close to the framework proposed by Brito {\it et al.}~\cite{brito2018}.

In real three-dimensional systems, the above rigidity transition is replaced by a continuous crossover, with a plateau shear modulus that appears smoothly below $T^*$, but with concave, square-root like dependence on temperature, qualitatively similar to that predicted by MCT. This was actually considered as one of the early success of MCT \cite{mct1,mct2}.

\end{document}